\documentclass[prd,twocolumn,showpacs,showkeys]{revtex4}
\usepackage{caption}
\usepackage{amsmath}
\usepackage{graphicx}
\usepackage{epsfig}
\usepackage{dcolumn}
\usepackage{bm}
\usepackage{slashed}
\usepackage{amsfonts}
\usepackage{amssymb}

\input{tcilatex}
\begin{document}

\title{When electric charge becomes also magnetic}
\author{T. C. Adorno$^{1}$}
\email{tadorno@usp.br; tg.adorno@gmail.com}
\author{D. M. Gitman$^{1,2,3}$}
\email{gitman@dfn.if.usp.br}
\author{A. E. Shabad$^{2,3}$}
\email{shabad@lpi.ru}
\affiliation{$^{1}$\textsl{Instituto de F\'{\i}sica, Universidade de S\~{a}o Paulo, CEP
05508-090, S\~{a}o Paulo, S. P., Brazil}\\
$^{2}$\textsl{P. N. Lebedev Physics Institute, Leninsky Prospekt 53, Moscow
\ 117924, Russia} \\
$^{3}$\textsl{Tomsk State University, Lenin Prospekt 36, Tomsk 634050, Russia%
} }

\begin{abstract}
In nonlinear electrodynamics, QED included, we find a static solution to the
field equations with an electric charge as its source, which is comprised of
homogeneous parallel magnetic and electric fields, and a radial
spherically-nonsymmetric long-range magnetic field, whose magnetic charge is
proportional to the electric charge and also depends on the homogeneous
component of the solution.
\end{abstract}

\keywords{magnetic monopole, nonlinear electrodynamics, magnetoelectric
effect in vacuum}
\pacs{11.10.Lm, 14.80.Hv, 42.50.Pq}
\maketitle

\section{Introduction}

Magneto-electric effect in materials is well-known theoretically and
experimentally \cite{LL}. It has a variety of manifestations, where
different schemes of interaction within the material lead either to linear
or nonlinear dependence of magnetization of an applied electric field, or
reciprocally, dependence of electrization on an applied magnetic field (see
reviews in \cite{obsor}). Calculation of the corresponding coefficients from
the first principles is available \cite{Mostovoy} in the linear case.

In the vacuum the (nonlinear) magnetoelectric effect was reported within
quantum electrodynamics (QED) in \cite{GitSha2012}, \cite{AdoShaGit2014},
and within noncommutative classical electrodynamics in \cite{AdoGitShaVas}
(There is an earlier indication of the magneto-electric effect in
noncommutative electrodynamics in Ref. \cite{Stern}. The magnetic solution
proposed there is, however, inadmissibly singular, see \cite{AdoGitShaVas}
for comments.) In these references the nonuniform magnetic field, associated
with a static electric charge is that of a magnetic dipole, whose magnetic
moment is quadratic in the charge. Here we find a magnetic field with its
source being an equivalent magnetic charge linearly related to an applied
static electric charge.

To be more precise, in a parity-conserving nonlinear electrodynamics of the
vacuum, especially in QED, we demonstrate the existence of a static field
configuration that possesses a magnetic charge. The magnetic field carrying
the magnetic charge is long-range in the sense that it decreases with the
distance $r=|\mathbf{x}|$\ from the electric charge, which produces it, as $%
r^{-2},$ in contrast to the magnetic dipole field decreasing as $r^{-3}$
from its center. The magnetic lines of force are directed along the
radius-vector $\mathbf{x.}\ $The magnetic nonuniform part of the solution is
necessarily accompanied by uniform constant electric and magnetic fields \ $%
\overline{\mathbf{E}}$\ and \ $\overline{\mathbf{B}}$\ \ (taken as parallel
to each other in our consideration), being axial-symmetric (i.e.,
azimuth-depending) relative to the axis specified by the common direction of
the uniform part of the solution. The magnetic charge defined as the surface
integral of the magnetic flux around the electric charge depends on these
constant field components, which are arbitrary, and it is proportional to
the electric charge and to the pseudoscalar $\overline{\mathbf{E}}\cdot 
\overline{\mathbf{B}}\ $. When the electric charge is pointlike, the
magnetic charge carried by it is pointlike, too. In other words, an electric
monopole is also a magnetic monopole. The vector-potential of the
axial-symmetric solution depends also on a choice of the angular boundary.
Its vector-potential is found to be singular, depending on a gauge, on
either of the two half-axes (Dirac string) drawn through the charge parallel
or antiparallel to the accompanying constant fields. For a very special
choice of the boundary, generally depending on the accompanying fields, the
Dirac string disappears leaving the axial-symmetric radial solution
magnetically neutral.

In this note we are paving the most straightforward way to a
magnetically-charged solution of nonlinear Maxwell equations by omitting all
the terms that may be under certain circumstances considered as inessential
for the existence of the solution proposed. We take the nonlinearity into
account in its simplest manifestation by keeping only the lowest nontrivial,
third power of the fields in these equations, while the corresponding
nonlinear self-coupling constant is considered to be small. When referring
to a nonlinear electrodynamics we mostly keep in mind the nonlinearity of
the Maxwell equations of QED stemming from the quantum phenomenon of
selfinteraction between electromagnetic fields. The approach, however,
remains the same for any nonlinear electrodynamics with a local Lagrangian
and also can be readily extended to include a parity-nonconserving
contribution that may originate from weak interactions.

\section{Nonlinear Maxwell equations}

Let the Lagrangian of a nonlinear theory be a function of the field
invariants $\mathfrak{F}=\left( B^{2}-E^{2}\right) /2,$ $\mathfrak{G}=-%
\mathbf{E}\cdot \mathbf{B}$, and let it depend on the space-time coordinate $%
x^{\mu }$ only through the fields and not contain their space-time
derivatives 
\begin{equation*}
L=-\mathfrak{F}(x)+\mathcal{L}(\mathfrak{F(}x\mathfrak{)},\mathfrak{G}\left(
x\right) ),
\end{equation*}%
which implies that the action $S=\dint L(x^{\prime })d^{4}x^{\prime }$ is a
local functional. The first term here is the Lagrangian of the standard
linear classical electrodynamics. The principle of correspondence with the
classical theory requires that the nonlinear addition $\mathcal{L}$ should
not contain any correction to it in the weak-field\ limit $B\rightarrow 0,$ $%
E\rightarrow 0$. Therefore it will be accepted that first derivative of $%
\mathcal{L}$ disappears when taken at zero values of the field invariants: $%
\mathcal{L}_{\mathfrak{F}}=\left. \frac{\partial \mathfrak{L}}{\partial 
\mathfrak{F}}\right\vert _{\mathfrak{F}=\mathfrak{G}=0}=0$.\ In a theory
with parity conservation, such as QED , the Lagrangian may depend only on
even powers of the pseudoscalar $\mathfrak{G}$. Hence, we shall accept that $%
\mathcal{L}_{\mathfrak{G}}=\left. \frac{\partial \mathfrak{L}}{\partial 
\mathfrak{G}}\right\vert _{\mathfrak{F}=\mathfrak{G}=0}=0,$ $\mathcal{L}_{%
\mathfrak{FG}}=\left. \frac{\partial ^{2}\mathfrak{L}}{\partial \mathfrak{F}%
\partial \mathfrak{G}}\right\vert _{\mathfrak{F}=\mathfrak{G}=0}=0$. These
quantities may be kept when needed. Under $\mathcal{L}$ we shall mostly mean
the effective Lagrangian of QED in the local (infrared) limit, given in the
one-loop approximation by the so-called Euler-Heisenberg Lagrangian \cite%
{BerLifPit}. Our approach covers, however, other nonlinear local Lagrangians
irrespective of their origin. The nonlinear static Maxwell equations
generated via the minimum action principle $\frac{\delta S}{\delta A_{\mu
}(x)}=j^{\mu }(x),$ where $A_{\mu }(x)$ is the vector-potential, with the
Lagrangian $\mathcal{L}$ truncated at the second power of its Taylor
expansion in the field invariants -- after the time-dependence has been
dropped from their form derived in \cite{CosGitSha2013} -- read%
\begin{equation}
\boldsymbol{\nabla }\cdot \mathbf{E}\left( \mathbf{x}\right) =j_{0}+j_{0}^{%
\text{nl}}\,,\text{ \ }\left[ \boldsymbol{\nabla }\times \mathbf{B}\left( 
\mathbf{x}\right) \right] =\mathbf{j}+\mathbf{j}^{\text{nl}}\,.  \label{ME}
\end{equation}%
Here $j_{\mu }$ are external current components, while the nonlinear current 
$j_{\mu }^{\text{nl}}$, cubic in the fields, is the one induced by the
electric $\mathbf{E}$ and magnetic $\mathbf{B}$ fields:%
\begin{eqnarray}
&&j_{0}^{\text{nl}}=\boldsymbol{\nabla }\cdot \left( \mathcal{L}_{\mathfrak{%
FF}}\mathfrak{F}\mathbf{E}+\mathcal{L}_{\mathfrak{GG}}\mathfrak{G}\mathbf{B}%
\right) \,,  \label{j0} \\
&&\mathbf{j}^{\text{nl}}=\left[ \boldsymbol{\nabla }\times (\mathcal{L}_{%
\mathfrak{FF}}\mathfrak{F}\mathbf{B}+\mathcal{L}_{\mathfrak{GG}}\mathfrak{G}%
\mathbf{E)}\,\right] .  \label{jvec}
\end{eqnarray}%
(It is understood that $\mathbf{E,\,B,\,\mathfrak{F}},$ and $\mathfrak{G}%
\mathbf{\ }$depend on $\mathbf{x}$. The $\nabla $%
\'{}%
s\ act on everything to the right of them). Here $\mathcal{L}_{\mathfrak{FF}}
$ and $\mathcal{L}_{\mathfrak{GG}}$ are time- and space-independent:%
\begin{equation}
\mathcal{L}_{\mathfrak{FF}}=\left. \frac{\partial ^{2}\mathfrak{L}}{\partial 
\mathfrak{F}^{2}}\right\vert _{\mathfrak{F}=\mathfrak{G}=0},\text{ \ }%
\mathcal{L}_{\mathfrak{GG}}=\left. \frac{\partial ^{2}\mathfrak{L}}{\partial 
\mathfrak{G}^{2}}\right\vert _{\mathfrak{F}=\mathfrak{G}=0}\,,
\label{L_FFL_GG}
\end{equation}%
taken at zero values of the fields, in other words the background, against
which the expansion of the Lagrangian has been developed is empty. Then, as
far as QED is concerned, $\mathcal{L}_{\mathfrak{FF}}$ and $\mathcal{L}_{%
\mathfrak{GG}}$ are quadratic with respect to the fine-structure constant 
\cite{BerLifPit} $\alpha =e^{2}/4\pi \simeq \left( 137\right) ^{-1},$%
\begin{equation}
\mathcal{L}_{\mathfrak{FF}}=\frac{4\alpha }{45\pi }\left( \frac{e}{m^{2}}%
\right) ^{2},\text{ \ }\mathcal{L}_{\mathfrak{GG}}=\frac{7\alpha }{45\pi }%
\left( \frac{e}{m^{2}}\right) ^{2}\,,  \label{BLP}
\end{equation}%
where $m$ and $e$ are electron mass and charge, respectively.

Equations (\ref{ME}) should be completed with the other pair of static
Maxwell equations%
\begin{equation}
\left[ \boldsymbol{\nabla }\times \mathbf{E}\left( \mathbf{x}\right) \right]
=0\,,\text{\ \ }\boldsymbol{\nabla }\cdot \mathbf{B}\left( \mathbf{x}\right)
=0\,,  \label{ME1}
\end{equation}%
which are intact to the nonlinearity as long as the fields are given by a
4-vector-potential.

\section{Magnetic solution for the field of electric charge}

In the present approach the electric field will be that produced by a
spherically-symmetric external charge $j_{0}\neq 0$. The Maxwell equations (%
\ref{ME}) will be treated perturbatively with respect to the small
selfcoupling contained in the coefficients $\mathcal{L}_{\mathfrak{FF}}$ and 
$\mathcal{L}_{\mathfrak{GG}}$. For this reason, the nonlinearly induced
charge density $j_{0}^{\text{nl}}$ (\ref{j0}) is neglected, $j_{0}^{\text{nl}%
}=0,$ as giving rise only to higher-order contribution. The magnetic field
will not be supported by any external source $\mathbf{j}$: the latter will
be kept equal to zero throughout, $\mathbf{j}=0$. The only source of the
magnetic field will be the current $\mathbf{j}^{\text{nl}},$ Eq.(\ref{jvec}%
), formed by the electric and magnetic fields themselves.

An additional approximation adopted in solving the nonlinear Maxwell
equations (\ref{ME}) is stemming from the fact that constant fields $\mathbf{%
E=}$ $\overline{{\mathbf{E}}}=$const${\mathbf{,}}$ $B=|\overline{\mathbf{B}}%
|=$const. identically satisfy the Maxwell equations (\ref{ME}) with no
external currents, $j_{0}=\mathbf{j}=0$, needed to support them, as it is
immediately seen from equations (\ref{ME}), (\ref{j0}), (\ref{jvec}), (\ref%
{ME1}). This is an approximation-independent manifestation of the gauge
invariance,\emph{\ }since the effective Lagrangian depends only on the field
strength, what makes any constant field a solution. We shall be solving the
second equation in (\ref{ME}) together with the second equation in (\ref{ME1}%
) for the magnetic deviation $\delta \mathbf{B}(\mathbf{x})=\overline{%
\mathbf{B}}-\mathbf{B}\left( \mathbf{x}\right) $ considered to be small as
compared to its constant part $\delta \mathbf{B}$ $\mathbf{\ll }$ $\overline{%
\mathbf{B}}.$ As for the electric field, its deviation from the constant
field $\delta \mathbf{E}(\mathbf{x})=\overline{\mathbf{E}}-\mathbf{E}\left( 
\mathbf{x}\right) $ will be taken, in correspondence with the above-said, as
the field produced by a charge via the standard linear equations $%
\boldsymbol{\nabla }\cdot \mathbf{E}\left( \mathbf{x}\right) =j_{0},$ $\left[
\boldsymbol{\nabla }\times \mathbf{E}\left( \mathbf{x}\right) \right] =0$
without nonlinearity. We also assume that $\delta \mathbf{E\ll }$ $\overline{%
\mathbf{E}}.$ We configure our solution with the fields $\overline{\mathbf{B}%
}$ and $\overline{\mathbf{E}}$ parallel or antiparallel to each other (in
the Lorentz frame, where the electric charge is at rest). Their arbitrary
common direction is presented by an unit (pseudo)vector $\boldsymbol{\mu }=%
\frac{\overline{\mathbf{B}}}{B}\,,$ \ $|\boldsymbol{\mu }\mathbf{|}=1$.

We shall now handle the second equation in (\ref{ME}) with $\mathbf{j}=0$,
which may be equivalently written as%
\begin{equation*}
(1-\mathfrak{F}\mathcal{L}_{\mathfrak{FF}})\left[ \boldsymbol{\nabla }\times 
\mathbf{B}\right] +\mathcal{L}_{\mathfrak{FF}}\left[ \mathbf{B}\times 
\boldsymbol{\nabla }\right] \mathfrak{F}-\mathcal{L}_{\mathfrak{GG}}\left[ 
\mathbf{E}\times \boldsymbol{\nabla }\right] \mathfrak{G}=0\,,
\end{equation*}%
because $\left[ \boldsymbol{\nabla }\times \mathbf{E}\right] =0$ due to (\ref%
{ME1}). Omitting the deviations squared and neglecting\emph{\ }$\overline{%
\mathfrak{F}}\mathcal{L}_{\mathfrak{FF}}\ll 1$\emph{,} this becomes the
linearized equation for $\delta \mathbf{B}$%
\begin{eqnarray}
\left[ \boldsymbol{\nabla }\times \delta \mathbf{B}\right]  &=&\mathcal{L}_{%
\mathfrak{FF}}\left[ \overline{\mathbf{B}}\times \boldsymbol{\nabla }\right]
\left( \overline{\mathbf{E}}\cdot \delta \mathbf{E}-\overline{\mathbf{B}}%
\cdot \delta \mathbf{B}\right)   \notag \\
&-&\mathcal{L}_{\mathfrak{GG}}\left[ \overline{\mathbf{E}}\times \boldsymbol{%
\nabla }\right] \left( \overline{\mathbf{B}}\cdot \delta \mathbf{E}+%
\overline{\mathbf{E}}\cdot \delta \mathbf{B}\right) \,.  \label{smallper}
\end{eqnarray}

We shall seek for solutions to equation (\ref{smallper}) in the class of
axial-symmetric magnetic fields directed in the same way as the
radius-vector, $\delta \mathbf{\mathbf{B=x}}b(r,z),$ where $r=|\mathbf{x}|,$
and $z=(\boldsymbol{\mu }\cdot \mathbf{x})=r\cos \theta $ is the coordinate
component along the vector $\boldsymbol{\mu }$. Within this class, the
second equation in (\ref{ME1}) can be satisfied only if the representation 
\begin{equation}
\delta \mathbf{\mathbf{B=x}}\frac{1}{r^{3}}f\left( \frac{z}{r}\right) \,,
\label{ansatzz}
\end{equation}%
is true (everywhere except for the singularity in $\mathbf{\mathbf{x=0}}$).

Out of all central-symmetric electric fields of the form $\delta \mathbf{E=x}%
\mathcal{E}(r)$, Eq. (\ref{smallper}) is only compatible with (\ref{ansatzz}%
) provided that the field is Coulombic, $\delta \mathbf{E}=\frac{c\mathbf{x}%
}{r^{3}},$ with $c$ being an arbitrary constant. But this is just the
solution to the first equation (\ref{ME}) with $j_{0}^{\text{nl}}$ set equal
to zero, as argued above, and with $c$ chosen as $c=\frac{q}{4\pi },$ where $%
q$ is the charge, $q=\dint j_{0}\left( \mathbf{x}\right) d^{3}\mathbf{x.}$
Here the integration run over the volume occupied by the charge centered in
the origin. (When the charge is distributed in a spherical-symmetric way
over a sphere $r=R$ or it is pointlike, $R=0,$ it should be understood that
we are working outside the charge, $r>R$. If the charge is not spherically
symmetric, our equations are valid far from the region occupied by it.) Then
the ansatz (\ref{ansatzz}) turns Eq. (\ref{smallper}) to the first-order
linear inhomogeneous differential equation for the function $f\left( \zeta
\right) ,$ $\zeta =\frac{z}{r}=\cos \theta $ (after cancellation of the
overall vector factor$[\boldsymbol{\mu }\times \mathbf{x}]/r^{4}$)%
\begin{align}
& (1+b\zeta ^{2})f^{\prime }+3bf\zeta +g\zeta =0,  \label{diffeq1} \\
& g=\frac{3q}{4\pi }\overline{\mathfrak{G}}\left( \mathcal{L}_{\mathfrak{GG}%
}-\mathcal{L}_{\mathfrak{FF}}\right) ,\text{ }b=-\mathcal{L}_{\mathfrak{GG}}%
\bar{E}^{2}-\mathcal{L}_{\mathfrak{FF}}\bar{B}^{2},  \label{g}
\end{align}%
where $\overline{\mathfrak{G}}=-\overline{\mathbf{B}}\cdot \overline{\mathbf{%
E}}.$ The general solution of (\ref{diffeq1})\ is%
\begin{equation}
f(\zeta )=\left( \frac{1+b\zeta _{0}^{2}}{1+b\zeta ^{2}}\right) ^{\frac{3}{2}%
}\left[ f(\zeta _{0})+\frac{g}{3b}\right] -\frac{g}{3b}\,.  \label{general}
\end{equation}%
Here $\zeta _{0}$ is the asimuth point, where a boundary $f(\zeta _{0})$
value for the solution is to be set. The singularity at $\zeta ^{2}=-1/b$
may lie within the physical interval $\zeta ^{2}\leq 1$ only for
sufficiently large fields $\bar{E},\bar{B}$, which are beyond the scope of
the present truncated approximation.

One can see that $f(\zeta )=-\frac{g}{3b}$ is the $\zeta $-independent
solution in immediate agreement with (\ref{diffeq1}) with $f^{\prime }=0$.
When substituted to (\ref{ansatzz}), it would give rise to a magnetic
monopole field in its standard center-symmetric form. This solution,
however, should rather be disregarded, since it falls beyond the accuracy of
the adopted approximation, as not being small in proportion to the smallness
of nonlinearity. It fades away, as being displaced to infinity, in the limit 
$b\rightarrow 0$. We shall concentrate in angle-dependent solutions, free of
this disadvantage. We are interested only in such solutions of the
inhomogeneous equation (\ref{smallper}), which are due to the electric
charge $q$ and vanish when $g=q=0$. To separate them we impose zero boundary
condition $f(\zeta _{0})=0$ (that also excludes the above angle-independent
solution). With this condition satisfied, the solution still depends on the
point $\zeta _{0}$, where it is imposed, i. e. on the couple of the asimuth
directions $\theta _{0}=\arccos \zeta _{0}$ and $\pi -\arccos \zeta _{0},$
along which the magnetic lines of force are not emitted from the charge. To
fix the boundary, note that $\zeta _{0}$ is a constant pseudoscalar, and
there may be no other choice of it except zero, since there is no
pseudoscalar at our disposal, and we should not introduce it once we are
interested in the magnetic field produced only by the electric charge and by
no other magnetically charged sources. The point $\zeta _{0}=0,$ $\theta
_{0}=\frac{\pi }{2}$ is the only point invariant under the space reflection,
because it is mapped to itself, since $\theta _{0}=$ $\pi -\theta _{0},$
when $\theta _{0}=\frac{\pi }{2}.$ With this choice the magnetic field (\ref%
{ansatzz}) is%
\begin{equation}
\delta \mathbf{\mathbf{B=x}}\frac{1}{r^{3}}\frac{g}{3b}\left[ \left( 1+b\cos
^{2}\theta \right) ^{-\frac{3}{2}}-1\right] \,.  \label{ansatzz2}
\end{equation}%
Note that this expression, as well as all other expressions in the rest of
the paper, survives the interesting limit $b=0,$ which originates from
dropping the homogeneous terms $\overline{\mathfrak{F}}\mathcal{L}_{%
\mathfrak{FF}}\left[ \boldsymbol{\nabla }\times \delta \mathbf{B}\right] ,$ $%
\mathcal{L}_{\mathfrak{FF}}\left[ \overline{\mathbf{B}}\times \boldsymbol{%
\nabla }\right] (\overline{\mathbf{B}}\cdot \delta \mathbf{B)}$ and $-%
\mathcal{L}_{\mathfrak{GG}}\left[ \overline{\mathbf{E}}\times \mathbf{\nabla 
}\right] \left( \overline{\mathbf{E}}\cdot \delta \mathbf{B}\right) $ from
Eq. (\ref{smallper}) and corresponds to neglect of linear magnetization.

The magnetic charge $q_{M}$ is determined by integrating $\delta \mathbf{B}$
over the surface of a sphere with its radius large enough to embrace the
whole charge and not to violate the conditions $\delta \mathbf{B}$ $\mathbf{%
\ll }$ $\overline{\mathbf{B}}$ and $\delta \mathbf{E}$ $\mathbf{\ll }$ $%
\overline{\mathbf{E}}.$ The result is 
\begin{equation}
q_{M}=\frac{4\pi g}{3b}\left[ \left( 1+b\right) ^{-\frac{1}{2}}-1\right] \,.
\label{magcharge}
\end{equation}%
Hence, for the pointlike electric charge, its magnetic charge density is $%
\mathbf{\nabla }\cdot \delta \mathbf{B}=q_{M}\delta ^{3}\left( \mathbf{x}%
\right) .$ The magnetic charge (\ref{magcharge}) is proportional to the
electric charge $q$ via Eq. (\ref{g}).

\section{Dirac string}

The vector potential generating the field (\ref{ansatzz2}) via the relation $%
\delta \mathbf{B}=\left[ \mathbf{\nabla }\times \mathbf{A}\right] $ can be
taken in the form%
\begin{eqnarray}
&&\mathbf{A}=\frac{\mathbf{[}\boldsymbol{\mu }\times \mathbf{x]}}{r^{2}}%
\omega (\zeta )\,,  \label{A} \\
&&\omega (\zeta )=\frac{g}{3b}\frac{1}{\zeta ^{2}-1}\left[ \frac{\zeta }{%
\sqrt{1+b\zeta ^{2}}}-\frac{\widetilde{\zeta }}{\sqrt{1+b\widetilde{\zeta }%
^{2}}}+\widetilde{\zeta }-\zeta \right] \,.  \notag
\end{eqnarray}%
The latter function is subject to the condition $\omega (\widetilde{\zeta }%
)=0.$ Arbitrariness in choosing the constant $\widetilde{\zeta }$ is the
gauge arbitrariness. The potential (\ref{A}) is singular along the axis $%
\zeta ^{2}-1=0.$ Two special gauges, $\widetilde{\zeta }=\pm 1,$ however,
exist, with which the axial singularity is restricted only to the positive
half-axis (the lower sign) or to the negative half-axis (the upper sign)$.$
With these two choices $\omega (\zeta )$ becomes%
\begin{equation*}
\omega ^{\pm }(\zeta )=\frac{g}{3b}\frac{1}{\zeta ^{2}-1}\left[ \frac{\zeta 
}{\sqrt{1+b\zeta ^{2}}}\mp \frac{1}{\sqrt{1+b}}-\left( \zeta \mp 1\right) %
\right] .
\end{equation*}%
It is seen by expanding the first term in the brackets in powers of $\zeta
^{2}$ near the point $\zeta ^{2}=1$ that $\left( \zeta \mp 1\right) $
cancels the same factor in the denominator if the latter is represented as $%
\zeta ^{2}-1=\left( \zeta \mp 1\right) \left( \zeta \pm 1\right) .$ The
resulting singularity $\left( \zeta \pm 1\right) ^{-1}$ stretches along the
axis passing through the charge either parallel or antiparallel to the
common direction of the constant part of the solution, the fields $\overline{%
\mathbf{B}}$ and $\overline{\mathbf{E}}.$ When the modulus of the
vector-potential is calculated following (\ref{A}), the factor $|\boldsymbol{%
\mu }\times \mathbf{x}|=\sqrt{1-\zeta ^{2}}$ appears, but we are still left
with the singularity $\pm \sqrt{\frac{1\mp \zeta }{1\pm \zeta }}$ on either
of the two half-axes. This is the Dirac string \cite{Dirac}, whose direction
depends on a gauge, but cannot be eliminated by any choice of it.

It is worth noting that a special choice of the boundary point $\zeta
_{0}^{2}=\frac{\sqrt[3]{1+b}-1}{b}$ exists that eliminates the Dirac string
and simultaneously nullifies the magnetic charge. This value depends on the
fields $\overline{B}$ and $\overline{E},$ but in the limit $b\rightarrow 0$
it is just $\zeta _{0}=\frac{\pm 1}{\sqrt{3}}.$ One may think that
introducing the boundary $\zeta _{0}$ other than zero leads to an additional
magnetic charge that neutralizes the magnetic charge of the electric charge
and results in the radial, but spherically nonsymmetric magnetic field with
the incoming and outcoming magnetic fluxes compensating each other.

\section{Concluding remarks}

With the QED values (\ref{BLP}) and the choice of the boundary $\theta _{0}=%
\frac{\pi }{2},$ and also neglecting $b$ as compared to unity the magnetic
charge is%
\begin{equation*}
q_{M}=\text{ }q\frac{\alpha }{30\pi }\frac{\overline{E}}{E_{0}}\frac{%
\overline{B}}{B_{0}},
\end{equation*}%
where $E_{0}$ and $B_{0}$ are the characteristic QED values $B_{0}=\frac{%
m^{2}}{e}=4.4\cdot 10^{13}\unit{G}$, $\ E_{0}=\frac{m^{2}}{e}=1.3\cdot
10^{16}\unit{V}/\unit{cm}$ (in CGSE units). Therefore, not too close to the
electric charge, all conditions assumed in the course of derivation of the
present result, including the requirement that $\overline{\mathfrak{F}}$ and 
$\overline{\mathfrak{G}}$ be smaller than $\left( m^{2}/e\right) ^{2},$
needed to justify the truncation of the effective action, are met. (The
latter restriction seems to be only technical and may be overcome by
expanding the action against the constant field background to be kept in (%
\ref{L_FFL_GG}). In that instance also the field-depending derivative $%
\mathcal{L}_{\mathfrak{FG}}$ will contribute). So, for the
astrophysical-scale values $\overline{B}\ \sim \overline{E}\lesssim \frac{%
m^{2}}{e}$ the magnetic charge makes up the 8$\cdot 10^{-5}$-th part of its
electric charge value.

It is important that the coefficient between the electric and magnetic
charges is a pseudoscalar, $\overline{\mathfrak{G}}$, whose presence in the
solution is necessary for the magneto-electric effect described. Any
possible field-configurations carrying magneto-electric effects, different
from ours, must contain this only pseudoscalar, too.

Note that in the limit $\overline{\mathbf{E}}=0,$ solution (\ref{ansatzz2})\
disappears not to turn into the magnetic solution of Refs. \cite{GitSha2012}%
, \cite{AdoShaGit2014} (which is of a dipole shape \cite{AdoShaGit2014})
produced by an electric charge in a constant magnetic background, because
that solution does not belong to the class considered here.

It is well understood that equations of electromagnetism readily accept a
magnetic charge, with the \textquotedblleft only\textquotedblright\
reservation that the latter has been never found in nature \cite{Cabrera},
except as a quasiparticle in spin ice \cite{spin-ice} or a physical
imitation \cite{Mottonen}. Here we have demonstrated that an electric charge
is also a magnetic one if\emph{\ }accompanied by\emph{\ }(placed into)\emph{%
\ }a combination of constant and homogeneous parallel electric and magnetic
fields. Evidently, the restriction on the fields to be parallel is crucial
only for the method of derivation, but the condition that they are not
mutually perpendicular, $\overline{\mathbf{E}}\mathbf{\cdot }\overline{%
\mathbf{B}}$\textbf{\ }$\mathbf{\neq 0,}$ cannot be circumvented. This means
that the charge may move in the Lorentz frame, where the fields $\overline{%
\mathbf{E}}$ and $\overline{\mathbf{B}}$ are parallel, without stop being a
magnetic monopole.

Supported by FAPESP under grants 2013/00840-9, 2013/16592-4 and
2014/08970-1, by RFBR under Project 14-02-01171, and by the TSU
Competitiveness Improvement Program, by a grant from \textquotedblleft The
Tomsk State University D.I. Mendeleev Foundation Program\textquotedblright .


\end{document}